\documentclass[]{llncs}
\pdfoutput=1
\usepackage[caption=false]{subfig}

\usepackage[utf8]{inputenc}
\usepackage{algorithm}
\usepackage{algorithmic}
\usepackage{times}

\usepackage{boxedminipage}

\usepackage{listings}

\usepackage{minitoc}

\usepackage{ifpdf}

\ifpdf
\usepackage[pdftex]{graphicx}
\else
\usepackage{graphicx}
\fi
\title{Multiresolution Cube Estimators for Sensor Network Aggregate Queries}
\author{Alexandra Meliou\inst{1} \and Carlos Guestrin\inst{2} \and Joseph M. Hellerstein\inst{3}}
\institute{University of Washington \email{ameli@cs.washington.edu}\footnote{This work was conducted while affiliated with University of California Berkeley.}
\and Carnegie Mellon University \email{guestrin@cs.cmu.edu}
\and University of California Berkeley \email{hellerstein@cs.berkeley.edu}}

\begin{document}

\ifpdf
\DeclareGraphicsExtensions{.pdf, .jpg, .tif}
\else
\DeclareGraphicsExtensions{.eps, .jpg}
\fi

\maketitle

\begin{abstract}
	In this work we present in-network techniques to improve the efficiency of spatial aggregate queries. Such queries are very common in a sensornet setting, demanding more targeted techniques for their handling. Our approach constructs and maintains multi-resolution cube hierarchies inside the network, which can be constructed in a distributed fashion. In case of failures, recovery can also be performed with in-network decisions. In this paper we demonstrate how in-network cube hierarchies can be used to summarize sensor data, and how they can be exploited to improve the efficiency of spatial aggregate queries. We show that query plans over our cube summaries can be computed in polynomial time, and we present a PTIME algorithm that selects the minimum number of data requests that can compute the answer to a spatial query. We further extend our algorithm to handle optimization over multiple queries, which can also be done in polynomial time. We discuss enriching cube hierarchies with extra summary information, and present an algorithm for distributed cube construction. Finally we investigate node and area failures, and algorithms to recover query results.
\end{abstract}

\section{Introduction}
Sensing devices are now used in many practical applications that require monitoring of physical phenomena. The data generated by such applications poses new challenges to data management research, prompting recent research on wireless sensor networks to devote significant attention to query processing \cite{maddengehrke04}. Model-based data acquisition schemes \cite{deshpande04} use historical information to predict rough query answers from probabilistic models, and optimize node selection to meet the desired query accuracy. Various centralized approaches \cite{meliou06,meliou07} aim to optimize communication strategies for data gathering, assuming accuracy of a centrally maintained model, which makes them more prone to failures. 

\looseness -1
Other query specific methods promote in-network modeling and decisions \cite{meliou09}, but they are geared towards general \texttt{\small SELECT *} type queries that request values at different sensor locations. In this work we address \emph{aggregate} queries, i.e. queries that request a certain type of summary information over a group of sensor nodes. Their differences from \texttt{\small SELECT *} queries require us to explore new approaches for the optimization of aggregates. Since sensornets are commonly used for monitoring physical phenomena over a specific area, such aggregates are usually spatially constrained, e.g. what is the average temperature in the engine room. \emph{Spatial interest queries} define a region of interest as a selection criterion, and query results are computed based on datapoints within that region.

Localized interest naturally lends itself to in-network summarization schemes, like the hierarchical summaries used in \cite{meliou09}. In this paper we propose \emph{multiresolution cube hierarchies} as a way to efficiently summarize and query spatially restricted aggregate data. This focus is well suited to the sensornet setting due to the type of applications that these environments usually support. Our scheme generalizes data cubes \cite{gray97} to represent area aggregates at different resolutions over the network. The cube hierarchies store information at different granularities (resolutions) allowing them to be applicable to queries of various ranges of region sizes.
Detailed description of the data stored in the network cube hierarchies is given in Section~\ref{sec:multiresolution_cubes}. 

We further demonstrate how the multiresolution cubes can be used to efficiently answer spatial aggregate queries. Section~\ref{sec:query_optimization} presents a polynomial algorithm that selects the minimum number of datapoints required to construct the answer for a specific query, and Section~\ref{sub:multiple_queries} extends it to the case of multi-query optimization. Section~\ref{sec:extensions} discusses an alternative method of summarization that maintains richer summaries, and presents an algorithm that performs distributed construction of the cube hierarchies. 

\looseness -1
Multiresolution cubes are effective against node failures, which commonly occur in sensor network settings. Section~\ref{sec:handling_failures} discusses isolated node failures as well as area outages, and demonstrates how the missing data can be recovered from other locations in the cube. Finally, Section~\ref{sec:related_work} discusses related work and Section~\ref{sec:conclusions} touches on future directions.

\section{Multiresolution Network Cubes} 
\label{sec:multiresolution_cubes}
\looseness -1
We focus on a type of in-network data summaries that will provide a framework to efficiently respond to aggregate queries with spatial constraints. Such queries are common in many applications, and are of the form \texttt{\small SELECT avg(temperature) WHERE nodeID inRegion [(2,3),(5,9)]}.
We target distributive (e.g. {\small SUM}) and algebraic aggregates (e.g. {\small AVG}), and assume queries that specify arbitrary rectilinear planar regions as their selection criterion. We do not address holistic aggregates in this work.
Regions of interest are defined as a set of points on the plane, and can be generalized to 3 dimensions, but in this paper we focus on the 2-dimensional case. In the simplest case, the region of interest is rectangular, and can be defined by just 2 corners, e.g. upper left and lower right. Regions can be of arbitrary shape, but those can always be split into simpler rectangles. The corner points of the interest areas can be given as coordinates in the euclidean space. 

For the purposes of this paper we will assume sensing locations arranged on a 2-dimensional grid, and regions of interest defined over this grid. Grid topologies have been studied before in the literature \cite{dimakis05,xu06}. Note however that this assumption does not restrict the application domain to grid deployments. Grid locations do not need to correspond to actual sensor locations, as the grid can be an overlay over the actual network topology, which we briefly discuss in Section~\ref{sec:conclusions}.

\begin{figure}[tb]
	\vspace{-2mm}
	\centering
		\includegraphics[height=1.3in]{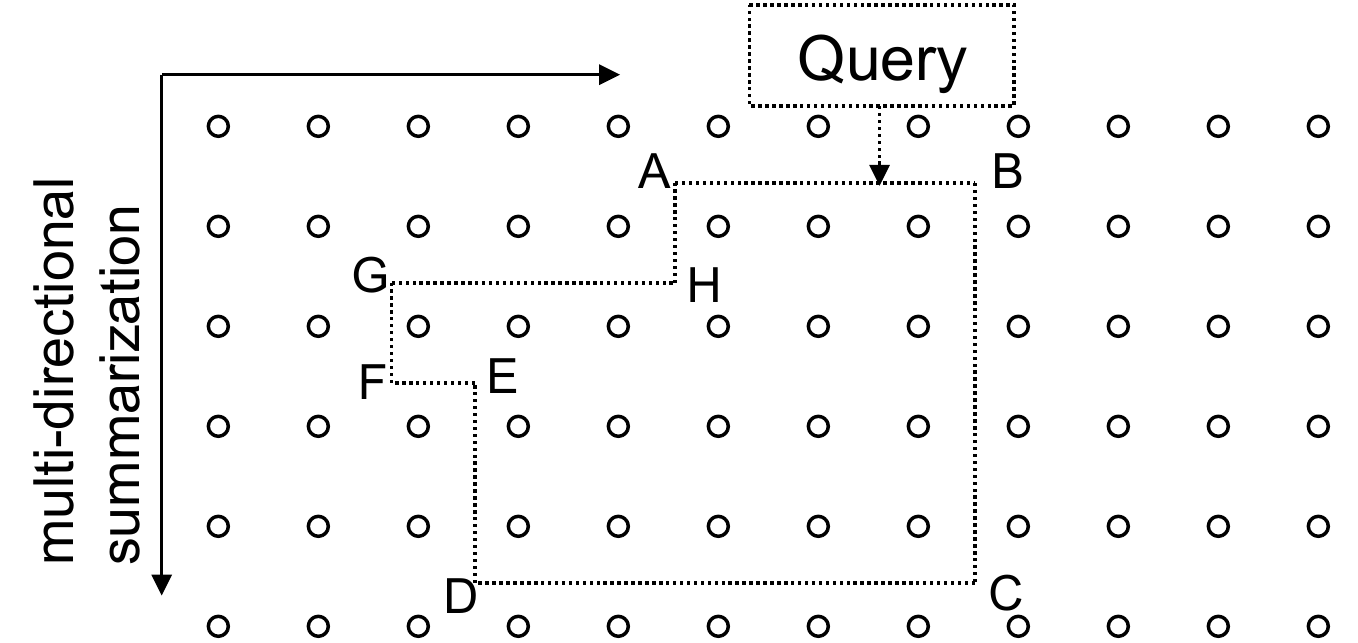}
	\caption{Spatial queries can define arbitrary rectilinear regions over the grid locations.}
	\label{fig:figs2_gridQuery}
\end{figure}

Queries define an area of interest over the grid specified by the locations of corner points, and request the computation of an aggregate value over the specified region. An example is shown in Figure~\ref{fig:figs2_gridQuery}. Containment of a grid location within the area of interest can be easily determined based on the region's point coordinates. Without loss of generality we will from now on focus on {\small SUM} as the aggregate function, but our scheme can be easily adapted to other distributive and algebraic aggregates, which can be expressed as a scalar function of distributive functions. 

In order to facilitate aggregate computation inside the interest region, we aim to preserve partial summary information within the network. The grid provides a natural setting for constructing cube summaries, as those can be computed along the grid directions, as in traditional data cubes. Without loss of generality we can assume that the top left corner of the grid is location $(0,0)$, with coordinates increasing from left to right and top to bottom, and that sum information is maintained along the increasing directions. These localized summaries can be stored at the corresponding grid locations and used to answer spatial queries without requiring the retrieval of data from all the corresponding nodes.

In order to accommodate queries of various region sizes and shapes, summarization can be performed at multiple granularity levels. The grid can be simply divided into rectangular cells of equal size, forming \emph{level-1} cubes where local summaries can be computed independently. The level-1 cubes cover the grid in a non-overlapping fashion and each summary is stored at a pre-specified cell location -- without loss of generality for this discussion we will assume that this location is the lower right corner of each cell.

\begin{figure}[b]
	\vspace{-2mm}
	\centering
		\includegraphics[height=1.5in]{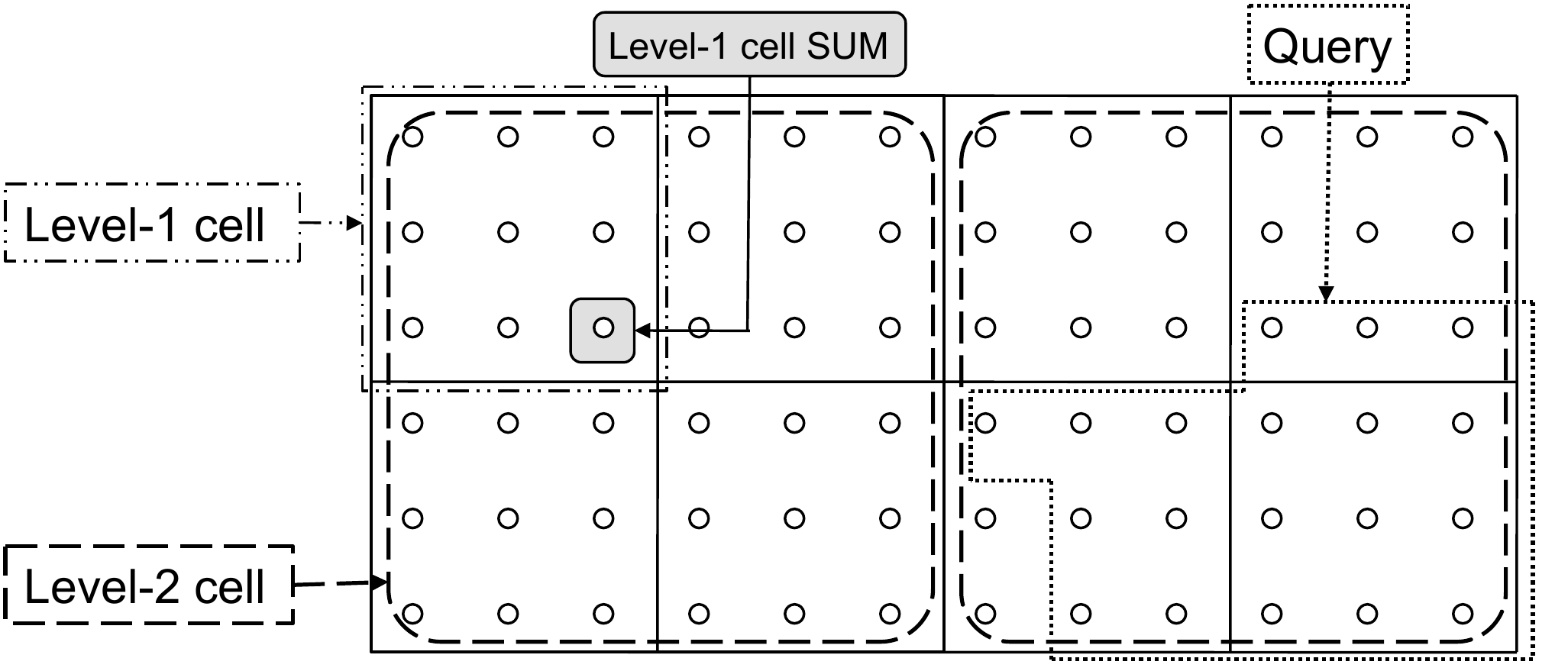}
	\caption{A two level cube hierarchy. Queries can span cells at different granularities.}
	\label{fig:figs2_cellsQ}
\end{figure}

Level-1 cells will now be treated as the new unit entities and get grouped to form level-2 cells. The process is repeated across multiple levels leading to a structure resembling a quad-tree.
A depiction of a simple two level hierarchy can be seen in Figure~\ref{fig:figs2_cellsQ}. In this example a level-1 cube cell contains the sum of the data from a $3\times 3$ area of the grid, and a level-2 cell sums over four level-1 cells. Note that a specific grid location may contain the sum data of more than one cell-level.
Due to the process of their construction, multiresolution cubes have the following properties: (a) cube cells of the same level do not intersect, (b) if cell $A$ of level $i$, and cell $B$ of level $j<i$ intersect, then $B\subset A$.

\subsection{Mapping Query Regions to Cube Cells} 
\label{sub:mapping_query_regions_to_cube_cells}

Cube summaries are stored in a distributed fashion in the network, at a specified location within the cell area that generated them, and can be used to compute aggregates over query specified regions without the need to access all the corresponding grid location. Query regions can be of arbitrary shape, spanning cells of different levels in the cube hierarchy, and our goal is to select the smallest number of cells that can reconstruct the requested aggregate. Since cube cells of the same level do not overlap, it is always most efficient to use the higher level cells that are contained in a region. 
An arbitrary query region can be decomposed into smaller rectangles based on the multi-resolution cube hierarchy. Then the region of interest becomes a collection of non-overlapping cells, whose data can be used to compute the aggregate over the whole region. Since cells of the same hierarchy level are not overlapping, the query area can be greedily split into cells in an optimal way, minimizing the number of cells that comprise it.

Algorithm~\ref{alg:greedyDivision} gives a sketch of the greedy division of a region into hierarchy cells. In the algorithm description, a corner of the interest region is \emph{convex} if the region is locally convex in the immediate area surrounding the corner. For example, in Figure~\ref{fig:figs2_gridQuery} the corners A, B, C, D, F, G of the query region are convex, while E and H are concave.
\begin{lemma}\footnote{Proofs can be found in the appendix.}\label{lem:greedyDiv}
	Algorithm~\ref{alg:greedyDivision} produces the minimum number hierarchy cells that exactly cover a given query region.
\end{lemma}
\begin{algorithm}\caption{\small Greedy Region Division (pseudocode)}\label{alg:greedyDivision}
	\footnotesize
	\begin{algorithmic}[1]
		\REPEAT
			\STATE Select a \emph{convex corner} $c$ of the region of interest $G$.
			\STATE Select $\max_k C_k$ ($C_k$ cell of level $k$) such that $C_k\subseteq G$ and $c\in C_k$.
			\STATE Extract $C_k$ from $G$.
		\UNTIL{$G=\emptyset$}
	\end{algorithmic}
\end{algorithm}

\section{Optimization of Spatial Aggregate Queries} 
\label{sec:query_optimization}
In the previous section we showed how a query region can be greedily mapped onto hierarchy cells, so that the minimum number of cells are used to cover the region. Algorithm~\ref{alg:greedyDivision} provides the optimal mapping of a query region to hierarchy cells, but this may not be the optimal strategy to compute a query aggregate.

\begin{figure}[tb]
    \centering
	\subfloat[]
    {\includegraphics[scale=0.32]{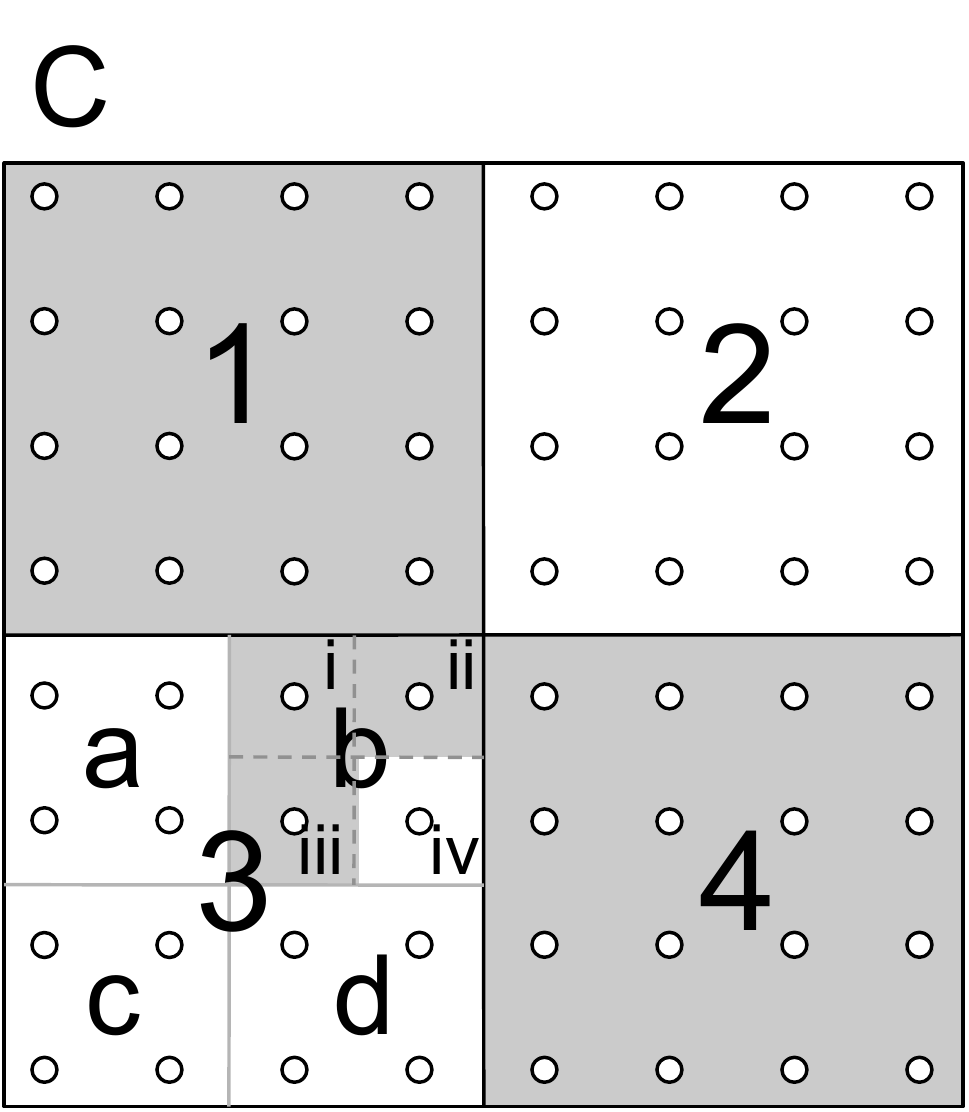}
        \label{fig:figs2_greyArea}}
	\hspace{4mm}
    \subfloat[]
    {\includegraphics[scale=0.4]{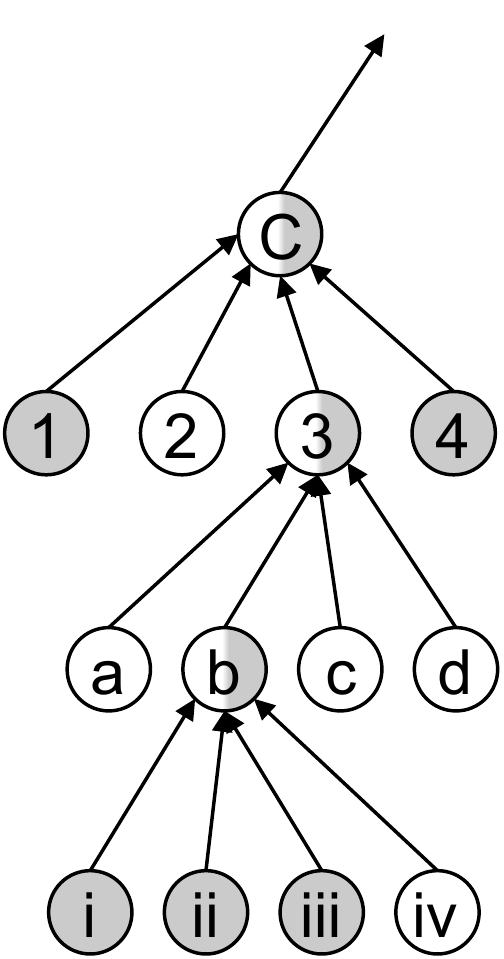}
        \label{fig:figs2_treeHierarchy}}
	\hspace{3mm}
	\subfloat[]
        {\includegraphics[scale=0.4]{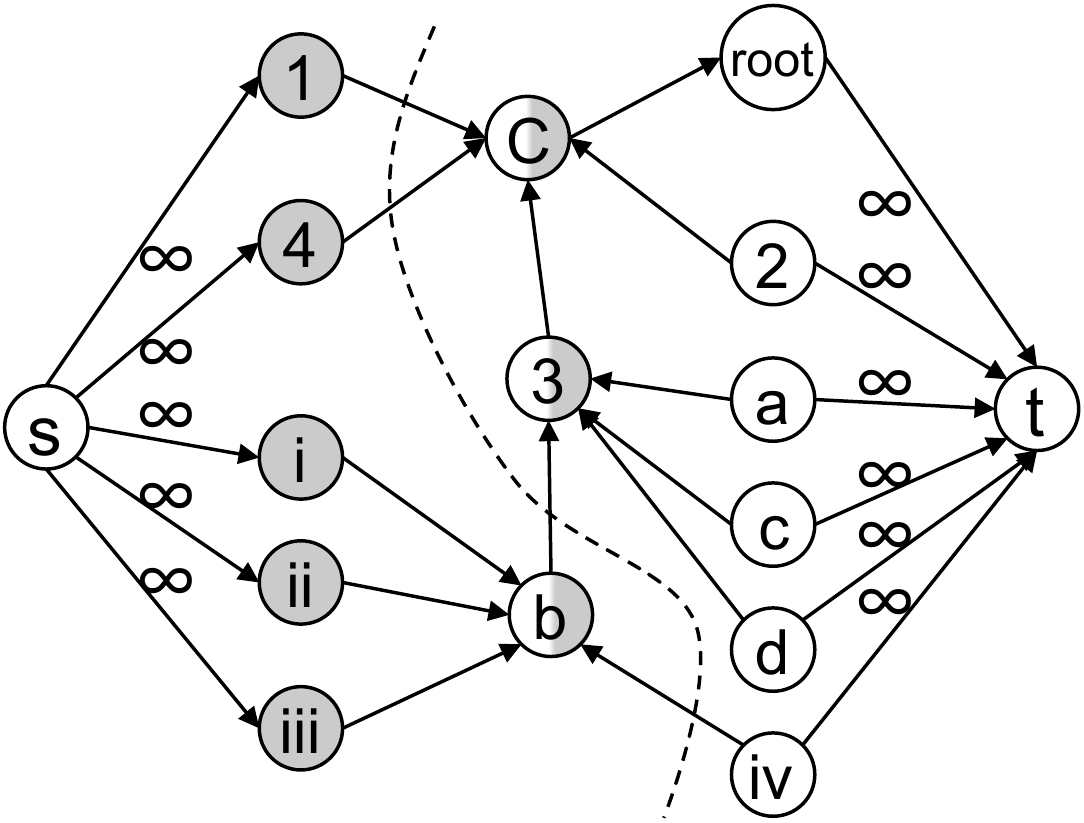}
		\label{fig:figs2_flow}}
	\caption{(a) The grey area depicts the area of interest of a query $G=\{1,4,i,ii,iii\}$ over the grid . (b) Multiresolution cube as a tree hierarchy. (c) Transformation to a max-flow problem}\label{fig:flowTransformation}
\end{figure}

In the example of Figure~\ref{fig:figs2_greyArea} cube C is further divided into 3 resolution levels, with some of the cell divisions displayed. Assume a spatial aggregate query over the shaded area $G$.
Algorithm~\ref{alg:greedyDivision} can map $G$ to the cube cells {\small $S={1,4,i,ii,iii}$}. The query aggregate can then obviously be computed using the summary values of cells in {\small $S$}: {\small $V(G)=V(1)+V(4)+V(i)+V(ii)+V(iii)$}, where {\small $V(x)$} refers to the aggregate value (sum) of cell or region $x$. The set {\small $S$} does not however provide a unique solution to the problem of aggregate reconstruction. {\small $V(G)$} can also be computed as {\small $V(C)-V(2)-V(a)-V(c)-V(d)-V(iv)$}, and there are numerous other possibilities. Most importantly, note that {\small $S$} does not provide the minimum solution either, in regards to the number of data points that need to be retrieved. In this example the smallest set of data points that can reconstruct {\small $V(G)$} is {\small ${1,4,b,iv}$: $V(G)=V(1)+V(4)+V(b)-V(iv)$}.

The problem that we want to solve is the following: given a multi-resolution cube hierarchy and a spatial aggregate query, select the minimum number of data points (cube aggregates) that are needed to compute the query answer. With this optimality criterion, the optimal query plan for a spatial aggregate query can be computed in polynomial time, through a polynomial reduction to a max flow problem. We will demonstrate the methodology with the example of Figure~\ref{fig:figs2_greyArea}.

A multi-resolution cube can be represented as a tree hierarchy, where every node represents a cube cell that contains all the cells in its subtree. Figure~\ref{fig:figs2_treeHierarchy} shows the tree representation of the example from Figure~\ref{fig:figs2_greyArea}. The grey-shaded nodes are cube cells completely contained in the query region $G$, white nodes do not overlap with $G$, and partially shaded nodes only partially overlap with $G$. Note that descendants of completely shaded or completely unshaded nodes are omitted, as their inclusion in the result would be suboptimal since a fully grey or fully white cell always dominates them. Edge direction simply denotes containment of cells across the different resolution levels.

The hierarchy can transformed into a flow problem as shown in Figure~\ref{fig:figs2_flow}. A source node $s$ is connected with infinite capacity edges to all the fully shaded nodes, and all white nodes with the inclusion of a root node as parent of the entire hierarchy are connected to a target node $t$, also with infinite capacity edges. The remaining edges are assigned unit capacity. This transformation conceptually partitions the nodes into those whose summary positively contributes to the query aggregate as they are contained in the interest region (shaded nodes), and those whose summary should not be included in the aggregate computation as they are outside the interest region (white nodes). Note that a \emph{cut} in the graph in Figure~\ref{fig:figs2_flow} simply assigns the partially shaded nodes into one of the two partitions. If a semi-shaded node is assigned to the grey partition (e.g $b$), then appropriate white nodes should be subtracted from the SUM (e.g. $iv$). Given an $s-t$ cut in the transformation graph with edge set $E$ that partitions the nodes into sets $S$ and $T$, with $s\in S$ and $t\in T$, then the total sum over the interest region is computed as 

{\small
$$
V(G)={\sum_{v\in S, u\in T\atop (v,u)\in E}V(v)} - {\sum_{v\in S, u\in T\atop (u,v)\in E}V(u)}
$$
}

The cut depicted in Figure~\ref{fig:figs2_flow} is actually the min-cut solution of size 4 for this graph, and corresponds to the summation {\small$V(G)=V(1)+V(4)+V(b)-V(iv)$}.
\begin{theorem}\label{thm:flowEquiv}
	The minimum number of data points from a multi-resolution cube hierarchy that are sufficient to answer a spatially constrained aggregate query is equal to the minimum $s-t$ cut in the corresponding flow graph. 
\end{theorem}
Theorem~\ref{thm:flowEquiv} provides a correspondence to a PTIME algorithm for computing the optimal set of data points in a cube hierarchy that can compute the answer to a spatial aggregate query. Using for example the Ford-Fulkerson algorithm to solve the corresponding max-flow problem we can select the optimal set of data points in $O(nf)$ where $f$ the maximum graph flow. In our case $f$ is bounded by $n$ which corresponds to the number of data points. 

\subsection{Dealing with Multiple Queries} 
\label{sub:multiple_queries}
\looseness -1
In the previous section we investigated optimal aggregate computation for a given spatially constrained query. However, when multiple queries run simultaneously in the system, optimizing them individually is not guaranteed to yield the overall optimal result.

\begin{figure}[tb]
    \centering
    \subfloat[]
    {\includegraphics[scale=0.45]{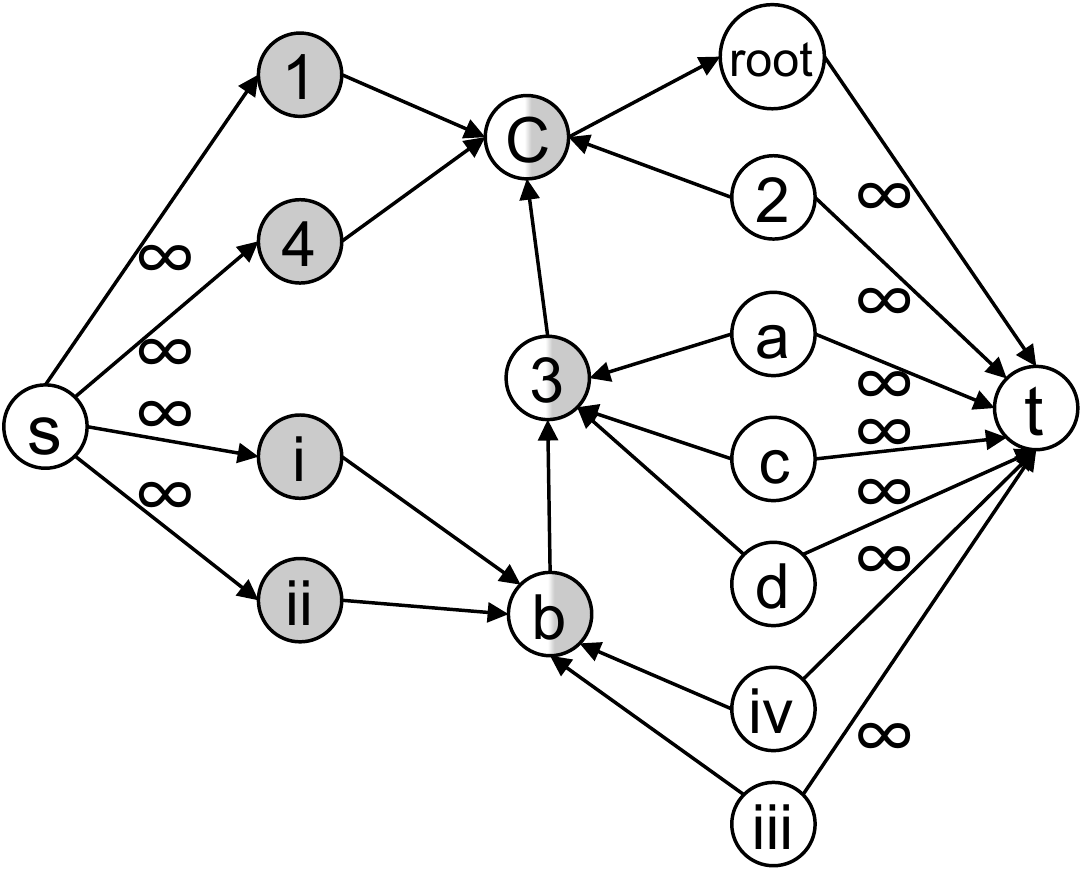}
        \label{fig:figs2_Q2merge}}
	\hspace{4mm}
	\subfloat[]
        {\includegraphics[scale=0.45]{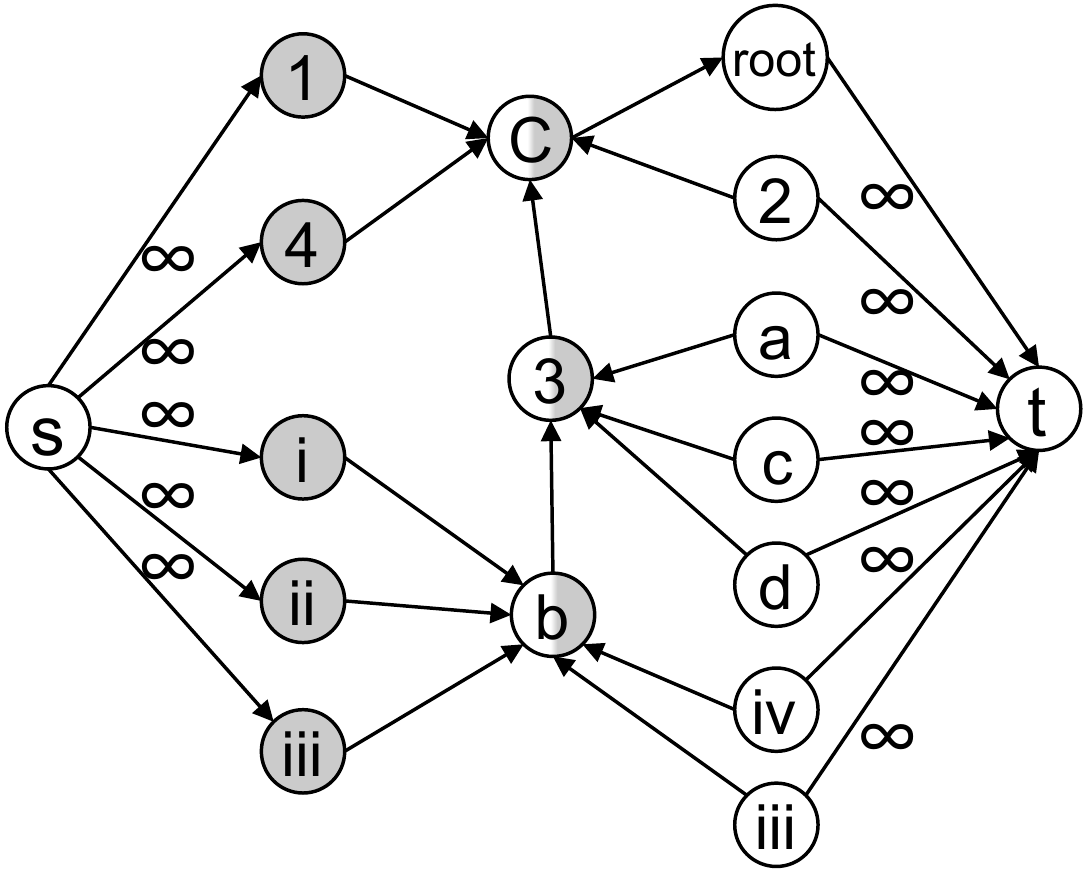}
		\label{fig:figs2_Q12combined2}}
	\caption{(a) Flow graph for query $Q_2$ (b) Combined flow graph for queries $Q_1$ and $Q_2$}\label{fig:combinedFlow}
\end{figure}

\looseness -1
Consider the cube hierarchy of Figure~\ref{fig:figs2_greyArea}, and {\small $Q_1$} a query with interest region the one depicted in the figure: {\small $G^{Q_1}=\{ 1,4,i,ii,iii\}$}. Also assume query {\small $Q_2$} for which {\small $G^{Q_2}=\{ 1,4,i,ii\}$}. The queries are mostly overlapping, apart from cell $iii$ which is included in {\small $Q_1$} but excluded in {\small $Q_2$}. The optimal set of data points that computes {\small $Q_1$} as seen in the previous section is $S_1=\{ 1,4,b,iv\}$. Similarly, computing the optimal set for {\small $Q_2$} using the flow graph of Figure~\ref{fig:figs2_Q2merge} yields set {\small $S_2=\{ 1,4,i,ii\}$}. This approach requires {\small $|S_1\cup S_2|=6$} data points to answer both queries.
Note however that this is not the best solution as we can compute both query results by retrieving just 5 elements: {\small $\{ 1,4,i,ii,iii\}$}.

The solution is to use a \emph{combined flow graph} by appropriately merging the individual flow graphs of all the queries we need to optimize for. In rough terms, the combined flow graph is a union of the individual graphs, creating replicas for nodes appearing in different groups (shaded/white/undetermined) in the base graphs. An example is given in Figure~\ref{fig:figs2_Q12combined2}, which is a combination of the graphs in Figures~\ref{fig:figs2_flow} and \ref{fig:figs2_Q2merge}. Note that in this example node $iii$ is shaded in one graph and white in the other, resulting in two instances, one white and one shaded, in the combined graph. Each of the original graphs is just a subgraph of the combined flow graph. 
\begin{proposition}\label{prop:combined}
	Two edges in the combined flow graph that correspond to the same data point cannot be part of the same minimum cut.
\end{proposition}
The combined graph contains all the information of the original graphs. A cut in the combined graph defines cuts in the original graphs, and therefore a cut in the combined graph corresponds to solutions for all queries that created it.

\section{Prefix-Sum Cubes} 
\label{sec:extensions}
In this section we will examine an alternative summarization scheme, which can enrich multi-resolution cubes with additional summary information. Up to this point we assumed a basic summarization approach that maintains the total sum of elements in each cube-cell. This method may under-utilize some of the grid locations, as cell summaries are kept in a subset of the nodes. 

\begin{figure}[bt]
	\centering
		\includegraphics[height=0.8in]{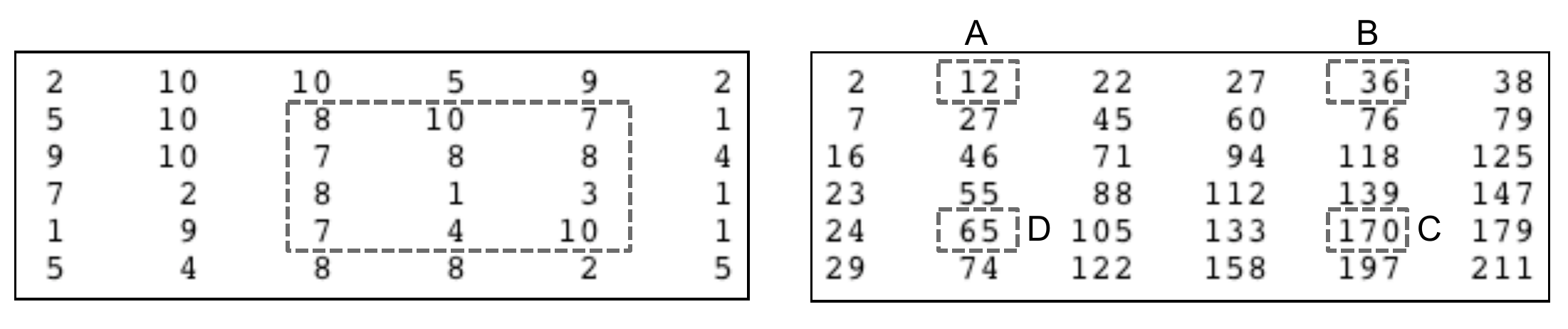}
	\caption{Example of Prefix-Sum. The sum within the denoted region can be computed from 4 data points: {\footnotesize$170+12-36-65=81$}}
	\label{fig:figs2_PS}
\end{figure}

A simple augmentation of the cell summaries that can better utilize the grid structure is the application of the \emph{prefix-sum} algorithm \cite{hillis86}. Prefix sum (PS) works on the grid by storing at location {\small $(i^*,j^*)$} the sum of the values from all locations where {\small $i\leq i^*$ and {\small $j\leq j^*$}}.
Figure~\ref{fig:figs2_PS} displays an example of the prefix sum algorithm: the first matrix contains the individual data, representing the values at each grid location, and the second matrix shows the result of the prefix-summation. 

PS values are stored in all nodes instead of just one node per cell. Again we can build multi-resolution hierarchies, by splitting the grid into cells at the lowest level, performing PS at each one, and reiterating at the next level with each cell as a new base element. An advantage of using prefix-sum is that it allows for finer granularities in the aggregate computation. Even a single level cube can allow the computation of the sum within a rectangular region using just the four corner points due to the way prefix-sums are computed. In the example of Figure~\ref{fig:figs2_PS} the region sum is equal to {\small $C+A-B-D$}.

\begin{proposition}\label{prop:numOfCorners}
	The total number of data points needed to calculate the sum within an arbitrary rectilinear region in a prefix-sum cube is the same as the number of its corners.
\end{proposition}

Cubes with PS information have more storage requirements, but can result in more efficient plans (fewer data points that need to be retrieved). A {\small $k\times k$} cube-cell stores {\small $2k^2-1$} different sums over the finer resolution level. Each of these data points refers to a rectangular region with upper left corner the upper left corner of the cell, and lower right corner the grid point {\small $(i,j)$} where this particular data point is stored. The data points are divided into three sets based on whether they overlap with the query region {\small $G$: $\forall s_i\in S_g$ $\{ s_i\}\cup G=G$, $\forall s_i\in S_w$ $\{ s_i\}\cap G=\emptyset$}, and {\small $\forall s_i\in S_u$ $\{ s_i\}\cap G\neq\emptyset$} and {\small $\{ s_i\}\cup G\supset G$}. 

We construct a ``re-colored'' set {\small $S_c$} from {\small $S_u$} as follows: {\small $\forall s_i\in S_u$} we select a subset {\small $S_w'\subset S_w$} such that {\small $s_i'=\{ s_i\}\setminus\bigcup_j\{ s_j\in S_w'\}$, $\{ s_i'\}\cup G=G$} and {\small $(\{ s_i\}\setminus\{ s_i'\})\cap G=\emptyset$}. We also assign {\small $cost(s_i')=|S_w'|+1$}. Finally {\small $S_c=\bigcup_i s_i'$}. We can now select the data points from the prefix sum cube that reconstruct the query with dynamic programming, using Algorithm~\ref{alg:PSquery}, where {\small $G$} is the grey area of interest, {\small $S=S_g+S_c$} and {\small $c=0$}.
\begin{algorithm}\caption{\small PSQuery(G,S,c)}\label{alg:PSquery}
	\footnotesize
	\begin{algorithmic}[1]
		\IF{$G=\emptyset$ or $S=\emptyset$}
			\STATE return c
		\ENDIF
		\FOR{s in S}
			\STATE i=i+1
			\STATE G'=G-s
			\STATE S'=\{$s\in S$ s.t. s contained in G\}
			\STATE c'=c+cost(s)
			\STATE A(i)=PSQuery(G',S',c')
		\ENDFOR
		\STATE return $\min(A(i))$
	\end{algorithmic}
\end{algorithm}

\subsection{Distributed Construction of Multi-resolution Cube Hierarchies} 
\label{sub:distributed_construction}
\looseness -1
Construction of the cube should be done in a distributed fashion, and should not be communication intensive. In this section we will present a distributed algorithm for the cube construction, which requires only a single transmission by every node. Our algorithm is designed to support the construction of prefix-sum hierarchies, but the additional prefix-sum values can be simply dropped to revert to the simple cube scheme discussed in Sec.~\ref{sec:multiresolution_cubes}.

\looseness -1
We assume that nodes know their location on the grid, and that a single packet has enough space for h values, where $h$ is the wanted height of the hierarchy. Every node will store up to $h$ values, depending on how many levels of the hierarchy it participates in.

\looseness -1
Given a set of fanouts $\{ F_i\}$ for the various hierarchy levels (number of cells that comprise higher level cells), a node can tell if it serves as a \emph{junction} for level $k$, i.e. the node that stores the sum of the current level-$k$ cell, if both its coordinates are divisible by $\prod_1^k{F_i}$. If a node is a junction for level $k$, it needs to forward its $k$-level sum to level $k+1$, but nothing to levels $>k+1$. A node adds its $k-1$ value, to the $k$-level message, thus adding to the $k$-level sum.
Also, a node knows based on its coordinates if it lies at the border of a level region. The specifics are given in Algorithm~\ref{alg:distrConstr}. With this scheme, every node broadcasts one packet, and receives 3. This allows the algorithm to scale very well, as the number of message transmissions and receipts is independent of the network size.

\vspace{3mm}
\begin{algorithm}\caption{\small Distributed Cube Construction (algorithm description)}\label{alg:distrConstr}
	\footnotesize
	\begin{algorithmic}[1]
		\STATE A node with coordinates (x,y) expects messages from nodes (x,y-1), (x-1,y) and (x-1,y-1). If one or more of these nodes don't exist ((x,y) is at the edge of the grid), then (x,y) proceeds without those messages.
		\STATE A packet is of the following form: $P=[F_1: value, F_2:value,\ldots,F_h:value]$.
		\STATE $F_0$ is defined as the local value at every node.
		\STATE Every node will store $k+1$ values, where $k$ is the level for which the node is a junction. By default all nodes are junctions for level 0.
		\STATE Upon receipt of the 3 packets $P_a$ from (x,y-1), $P_b$ from (x-1,y) and $P_c$ from (x-1,y-1), it computes $P(F_i)=P_a(F_i)+P_b(F_i)-P_c(F_i)$.
		\STATE If x-1 is divisible by $\prod_1^k{F_i}$, then node (x,y) sets $P_b(F_k)=P_c(F_k)=0$ before making the above computation. If y-1 is divisible by $\prod_1^k{F_i}$, then node (x,y) sets $P_a(F_k)=P_c(F_k)=0$ before making the above computation. 
		\STATE A k-level junction node stores level values up to level $k+1$, where for level $i$ the value is $P(F_i)+local(F_{i-1})$.
		\STATE The node builds a new packet $P$ with the $k+1$ values that it has stored. It also populates it with the values $P(F_{k+2}),\ldots P(F_h)$ as computed upon receipt.
	\end{algorithmic}
\end{algorithm}

\section{Handling Failures} 
\label{sec:handling_failures}
In the previous sections we discussed the structure and distributed construction of multi-resolution cube hierarchies, either with simple or prefix-sum summarization schemes, and developed algorithms that use these in-network estimators to efficiently answer spatially constrained queries. In this section we will discuss data recovery after node failures, which is a common occurrence in sensornet deployments. Our multi-resolution cubes implement data redundancy, making full recovery feasible in many cases of failures. Especially in the case of prefix-sum hierarchies, most node values can be reconstructed by simply querying 3 immediate neighbors.

The k-level value for a specific node is constructed as {\small $local(F_i) = P_a(F_i)+P_b(F_i)-P_c(F_i)+local(F_{i-1})$} (see Algorithm~\ref{alg:distrConstr}). Therefore, any node that is used as an $a$,$b$ or $c$-node in this equation (locations {\small$(x,y-1)$}, {\small $(x-1,y)$} and {\small$(x-1,y-1)$} respectively) can have its value reconstructed by querying the other 3 nodes in the local square.
The only complication arises with junction nodes. A junction node does propagate its value in the same fashion, but the receiving nodes do not store it but only forward it until it reaches the appropriate node for that hierarchy level. That means that if a node that serves as a junction for level $i$ fails, then we need to access data at a total distance of {\small$3F_i$} ({\small$F_i$} the current fan-out) from that node to reconstruct the missing value.

Another alternative would be to store at every node up to level {\small $k+2$} values instead of {\small $k+1$}. The construction scheme would not change, as that information is already sent around the network, but nodes would be required to store one extra value. That would allow the reconstruction of the value after a single failure with just  querying the 3 immediate neighbors. This implies a tradeoff between storage space and recovery capability, which can be further investigated in future work.

An important point to note is that the recovery logic is tied to the method of construction, which is performed with local planning and decisions. The location of data points that can reconstruct missing measurements can be determined locally, and therefore upon detection of failures, recovery can be performed with in-network decisions.  

\subsection{Area Failures} 
\label{sub:area_failures}
Isolated node failures can be resolved in a straightforward fashion using the values of the neighbors of the failing nodes.
Area failures introduce more complications, but they can still be addressed, and fully reconstructed in many cases. We identify area failures as those that affect an arbitrary number of consecutive nodes on the grid, and they may extend to different resolutions of cell levels.
Known areas of failure can easily be bypassed during data point selection using the transformation graph described in Section \ref{sec:query_optimization}, by setting the appropriate edge capacities to infinity. For example, if cell 4 failed, setting the capacity of edge $(4,C)$ to $\infty$, would force the algorithm to select a different solution.

Note that areas can fail in such a way that the failures cannot be bypassed for some queries. For the query of Figure~\ref{fig:figs2_greyArea}, if areas 2 and 4 fail, there is no way of retrieving the exact answer. Note that in this case there would be no cut in the flow graph with weight $<\infty$. It is interesting to note that it is possible to compute the total aggregate value in the combined $\{ 2,4\}$ area, but the query only intersects part of that region.

\begin{figure}[tb]
    \centering
    \subfloat[]
    {\includegraphics[scale=0.375]{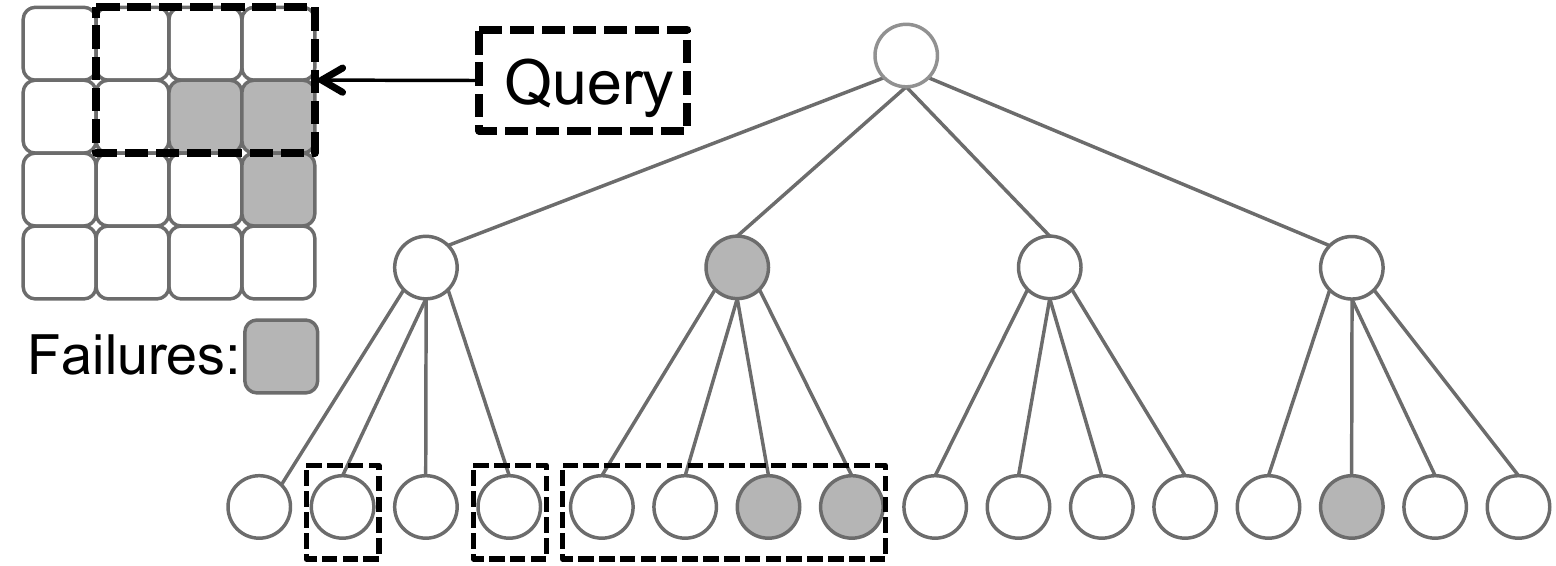}
        \label{fig:figs2_areaFail}}
	\hspace{0mm}
	\subfloat[]
        {\includegraphics[scale=0.375]{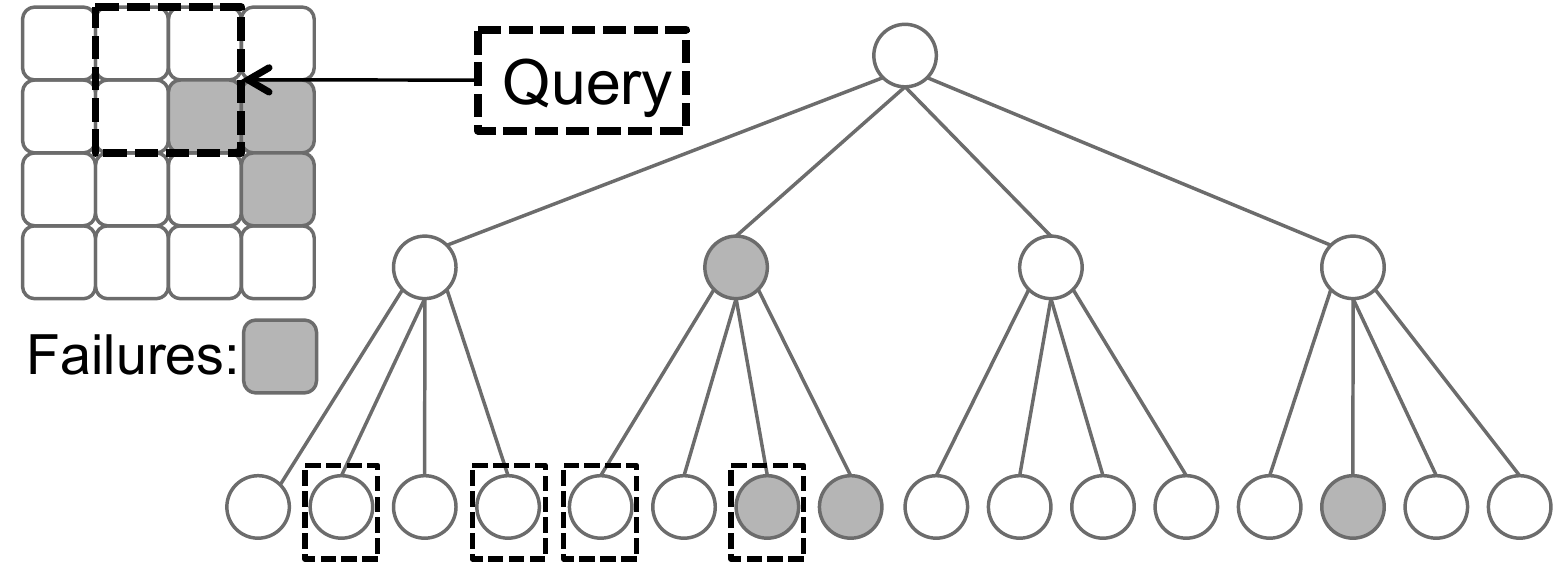}
		\label{fig:figs2_areaFail2}}
	\caption{(a) Example query that can be computed accurately (b) Example query that cannot be computed accurately}\label{fig:areaFailures}
\end{figure}

\looseness -1
Using the flow reduction, it can be therefore determined in polynomial time whether the query can be answered accurately or not. Figure~\ref{fig:figs2_areaFail} shows an example where the query value can be computed accurately. In the cases where computing the accurate answer is infeasible, a solution should be found that best approximates it. Without further knowledge of correlations or other information on the actual data, a reasonable approach is to assume uniformity over the values in the failed region and use it to estimate the aggregate in the portion of it that is requested by the query. To minimize inaccuracies due to the uniformity assumption, this estimation should be done in as small a portion of the failed region as possible, relative to the region of interest. An interesting issue for future work is to further investigate various approaches to construct estimates for non-recoverable regions. 

In the example of Figure~\ref{fig:figs2_areaFail2} the query intersects only $\frac{1}{3}$ of the failed region, and the exact answer cannot be retrieved. However, instead of estimating the sum of the overlapping region as $\frac{1}{3}$ of the full failed area, it is probably better to estimate it as $\frac{1}{2}$ of the top 2-cell portion of the failed region, which can be accurately reconstructed.

Algorithm~\ref{alg:regionRecovery} traverses a cube hierarchy like the ones in Figure~\ref{fig:areaFailures} bottom up, starting with an area A matching the query region that contains failures. If the sum over A cannot be computed, the algorithm traverses to higher levels and the area A gets augmented as necessary. The algorithm returns the exact answer if eventually A does not get augmented, otherwise returns an estimate based on the uniformity assumption, over the portion of the failed region that has been recovered.

\vspace{3mm}
\begin{algorithm}\caption{\small Region Recovery}\label{alg:regionRecovery}
	\footnotesize
	\begin{algorithmic}
		\STATE Start at leaf level failed regions $Q_A$. Initialize $V=0$ and $A=Q_A$;
		\WHILE{fail(currentNode)}
			\STATE level$--$;
			\IF{$fail(children(currentNode))>1$}
				\STATE $A=A\cup newLeafFailures$;
				\STATE $V=V-\sum{V(aliveLeavesInNewFailure)}$;
			\ENDIF
			\STATE $V=V-\sum{V(aliveChildren)}$;
			\IF{$!fail(currentNode)$}
				\STATE $V=V+V(currentNode)$;
			\ENDIF
		\ENDWHILE
		\IF{$A>Q_A$}
			\STATE region $Q_A$ cannot be fully recovered
		\ENDIF
		\STATE return $estimate=V\frac{Q_A}{A}$
	\end{algorithmic}
\end{algorithm}

\section{Related Work} 
\label{sec:related_work}
Spatial query processing has been extensively studied in centralized systems. The R-tree \cite{guttman84}, and its variants (\cite{beckmann90,sellis87}), is one family of index structures for spatial data. In the R-tree, each spatial data object is represented by a Minimum Bounding Rectangle containing a pointer to the object in the database. Non-leaf nodes store a MBR that contains the MBRs of all the children nodes. A query traverses the R-tree using ``containment'' and ``overlap'' checks to appropriately navigate through the structure. Spatial indexing is discussed more extensively in \cite{guting94}.

Because of resource limitations in sensor networks, building centralized indexes is often not practical. \cite{dermibas03} proposes a peer-tree, a distributed R-tree using peer-to-peer techniques, partitioning the sensor network into hierarchical, rectangle shaped clusters. The techniques include joins and splits of clusters, and the authors show how to use the structure to answer nearest neighbor queries.
The use of quad-trees in such setting is also natural and \cite{dermibas07} uses distributed quad-trees to support spatial querying. 

SPIX \cite{soheili05}, is a distributed spatial index which uses Minimum Bounding Areas in an R-tree like fashion. The spatial query processor on each sensor uses SPIX to bound the branches that do not lead to results, find a path to sensors that do have results to report, as well as aggregate data.

In the database community, a number of distributed and push-down based approaches have been proposed for aggregation (\cite{shatdal95,yan95}). However these assume a well connected, low-loss topology that is often unrealistic in sensor networks. TAG (\cite{madden02,madden02b}) performs in-network aggregation to reduce the amount of data that needs to be send over the network. TAG provides a simple declarative interface for aggregation, and it distributes and executes aggregation operators in the network, computing aggregates through data flow. Aggregation in adversarial settings is examined in \cite{garofalakis07}.

We use ideas from Data Cubes \cite{gray97} and simple summarization techniques \cite{hillis86}. The aR-trees \cite{Papadias:2001p449} introduce OLAP type aggregation indexes over spatial data, resembling our data cube hierarchies, but are not oriented to a distributive environment, making some construction and failure issues irrelevant.
Our approach can also relate to methods of distributed storage in sensor networks \cite{aly08,dimakis06}, where data get disseminated and encoded at different locations to improve resilience to failures. Our approach does not simply target failures, but also improves processing for spatial aggregates, which commonly occur in these settings.

\section{Conclusions and Future Directions} 
\label{sec:conclusions}
In this work we presented multi-resolution cube estimators that keep summary information distributed in the network, and facilitate the computation of spatial aggregate queries. We presented algorithms that compute query plans over the cube hierarchies in polynomial time, and showed how they can also accommodate multiple queries. We presented a distributed scalable algorithm that performs the cube construction and algorithms that perform failure recovery. We demonstrated that multi-resolution cubes are very resilient to node failures, and also behave well in larger region outages.

In this paper we focused on a grid topology, but it is an interesting problem to extend this approach as an overlay over any deployment. Using observations at the actual sensor locations and spatial models of the data distributions, we can infer the values at the grid locations. This would require us to adapt our algorithms to account for probabilistic data and the possible correlations across the grid locations. Especially in the case of prefix-sum cubes, which maintain summaries of overlapping areas, these correlations cannot be disregarded. Simple sum cubes are easier to extend as the summary regions do not intersect, making the problem less complex.

\bibliographystyle{splncs}
\bibliography{refer}

\begin{thebibliography}{10}

\bibitem{maddengehrke04}
Madden, S., Gehrke, J.:
\newblock Query processing in sensor networks.
\newblock Pervasive Computing \textbf{3}(1) (2004)

\bibitem{deshpande04}
Deshpande, A., Guestrin, C., Madden, S., Hellerstein, J., Hong, W.:
\newblock Model-driven data acquisition in sensor networks.
\newblock In: VLDB. (2004)

\bibitem{meliou06}
Meliou, A., Chu, D., Guestrin, C., Hellerstein, J., Hong, W.:
\newblock Data gathering tours in sensor networks.
\newblock In: IPSN. (2006)

\bibitem{meliou07}
Meliou, A., Krause, A., Guestrin, C., Hellerstein, J.M.:
\newblock Nonmyopic informative path planning in spatio-temporal models.
\newblock In: AAAI. (2007)

\bibitem{meliou09}
Meliou, A., Guestrin, C., Hellerstein, J.M.:
\newblock Approximating sensor network queries using in-network summaries.
\newblock In: Information Processing in Sensor Networks (IPSN). (2009)

\bibitem{gray97}
Gray, J., Chaudhuri, S., Bosworth, A., Layman, A., Reichart, D., Venkatrao, M.,
  Pellow, F., Pirahesh, H.:
\newblock Data cube: {A} relational aggregation operator generalizing group-by,
  cross-tab, and sub-totals.
\newblock J. Data Mining and Knowledge Discovery \textbf{1}(1) (1997)  29--53

\bibitem{dimakis05}
Dimakis, A.G., Prabhakaran, V., Ramchandran, K.:
\newblock Ubiquitous access to distributed data in large-scale sensor networks
  through decentralized erasure codes.
\newblock In: IPSN, Piscataway, NJ, USA, IEEE Press (2005)

\bibitem{xu06}
Xu, K., Takahara, G., Hassanein, H.:
\newblock On the robustness of grid-based deployment in wireless sensor
  networks.
\newblock In: IWCMC. (2006)  1183--1188

\bibitem{hillis86}
Hillis, W.D., Steele, Jr., G.L.:
\newblock Data parallel algorithms.
\newblock Commun. ACM \textbf{29}(12) (1986)  1170--1183

\bibitem{guttman84}
Guttman, A.:
\newblock {R}-trees: a dynamic index structure for spatial searching.
\newblock In: SIGMOD, ACM (1984)  47--57

\bibitem{beckmann90}
Beckmann, N., Kriegel, H.P., Schneider, R., Seeger, B.:
\newblock The {R}*-tree: an efficient and robust access method for points and
  rectangles.
\newblock In: SIGMOD, ACM (1990)  322--331

\bibitem{sellis87}
Sellis, T.K., Roussopoulos, N., Faloutsos, C.:
\newblock The {R}+-tree: A dynamic index for multi-dimensional objects.
\newblock In: VLDB. (1987)  507--518

\bibitem{guting94}
G\"{u}ting, R.H.:
\newblock An introduction to spatial database systems.
\newblock The VLDB Journal \textbf{3}(4) (1994)  357--399

\bibitem{dermibas03}
Demirbas, M., Ferhatosmanoglu, H.:
\newblock Peer-to-peer spatial queries in sensor networks.
\newblock In: P2P, IEEE Computer Society (2003) ~32

\bibitem{dermibas07}
Demirbas, M., Lu, X.:
\newblock Distributed quad-tree for spatial querying in wireless sensor
  networks.
\newblock ICC (2007)

\bibitem{soheili05}
Soheili, A., Kalogeraki, V., Gunopulos, D.:
\newblock Spatial queries in sensor networks.
\newblock In: GIS, ACM (2005)  61--70

\bibitem{shatdal95}
Shatdal, A., Naughton, J.F.:
\newblock Adaptive parallel aggregation algorithms.
\newblock SIGMOD Rec. \textbf{24}(2) (1995)  104--114

\bibitem{yan95}
Yan, W.P., Larson, P.A.:
\newblock Eager aggregation and lazy aggregation.
\newblock In: VLDB. (1995)  345--357

\bibitem{madden02}
Madden, S., Franklin, M.J., Hellerstein, J.M., Hong, W.:
\newblock {TAG}: a tiny aggregation service for ad-hoc sensor networks.
\newblock SIGOPS Oper. Syst. Rev. \textbf{36}(SI) (2002)  131--146

\bibitem{madden02b}
Madden, S., Szewczyk, R., Franklin, M.J., Culler, D.:
\newblock Supporting aggregate queries over ad-hoc wireless sensor networks.
\newblock In: WMCSA. (2002) ~49

\bibitem{garofalakis07}
Garofalakis, M.N., Hellerstein, J.M., Maniatis, P.:
\newblock Proof sketches: Verifiable in-network aggregation.
\newblock In: ICDE, IEEE (2007)  996--1005

\bibitem{Papadias:2001p449}
Papadias, D., Kalnis, P., Zhang, J., Tao, Y.:
\newblock Efficient {OLAP} operations in spatial data warehouses.
\newblock Advances in Spatial and Temporal Databases (2001)  443--459

\bibitem{aly08}
Aly, S.A., Kong, Z., Soljanin, E.:
\newblock Fountain codes based distributed storage algorithms for large-scale
  wireless sensor networks.
\newblock In: IPSN. (2008)  171--182

\bibitem{dimakis06}
Dimakis, A.G., Prabhakaran, V., Ramchandran, K.:
\newblock Decentralized erasure codes for distributed networked storage.
\newblock IEEE/ACM Trans. Netw. \textbf{14}(SI) (2006)  2809--2816

\end{thebibliography}

\newpage
\appendix
\section{Proofs} 
\label{sec:proofs}

\begin{proof}[Lemma~\ref{lem:greedyDiv}]
    Assume $S$ the set of hierarchy cells produced by Algorithm~\ref{alg:greedyDivision}, and $S'$ another set of hierarchy cells such that $S'$ also precisely covers the query region $R$, and $|S'|<|S|$. Then there must exist a grid location that is covered by a higher level cell $C_i$ in $S'$ than in $S$ ($C_j$). Then $C_j\subset C_i$. Since the algorithm always selects the highest level cells, that could only happen if an earlier greedy step had removed part of the region covered by $C_i$. Cells do not overlap unless one contains another. Given a corner $c$, if the algorithm selects $C_k$ when $C_k\subset C_i$ and $C_i$ is fully contained in the query region, that means that $c\in C_k$ and $c\not\in C_i$. That would mean that $c$ was not a convex corner, which is disallowed by the algorithm. Therefore $S$ is minimum.\qed
\end{proof}

\begin{proof}[Theorem~\ref{thm:flowEquiv}]
    An $s-t$ cut in the flow graph partitions the nodes into sets $S$ and $T$ with $s\in S$ and $t\in T$. Also assume $V_g$ the set of all grey nodes, $V_w$ the set of all white nodes plus the ``root'' node, and $V_u=V-V_g-V_w$ the rest ``unassigned'' nodes. Then $V_g\in S$ and $V_w\in T$ due to the infinite capacity edges. Note that $\forall v\in V_u$ there is a single outgoing edge $(v,v_o)$ and multiple incoming edges $(u_i,v)$. We also know from cell containment between different resolutions that $V(v)=\sum_i V(u_i)$.
    
    Assume a cut $\{ S,T\}$ such that $v\in T$. The cut would then contain a set of edges $E_1$ such that $(u_i,v)\in E_1\Leftrightarrow u_i\in S$ (Figure~\ref{fig:figs2_flip}).
    Then the query aggregate would be: $$V(G)=s_1+\sum_{(u_i,v)\in E_1} V(u_i)$$ where $s_1$ the sum of weights based on the rest of the edges, apart from $E_1$, that are in the cut.
    The cut $\{ S'=S\cup\{ v\},T'=T\setminus\{ v\}\}$ differs from $\{ S,T\}$ only in the inclusion of $v$. The query aggregate for the $\{S',T'\}$ cut would then be computed as $V(G)'=\sum_1+ V(v)-\sum_{(u_i,v)\in E_2} V(u_i)$. But as noted earlier, $\sum_{(u_i,v)\in E_1} V(u_i) = V(v)-\sum_{(u_i,v)\in E_2} V(u_i)$, therefore $V(G)'=V(G)$.
    \begin{figure}[ht]
        \addtolength{\belowcaptionskip}{3mm}
        \centering
            \includegraphics[scale=0.5]{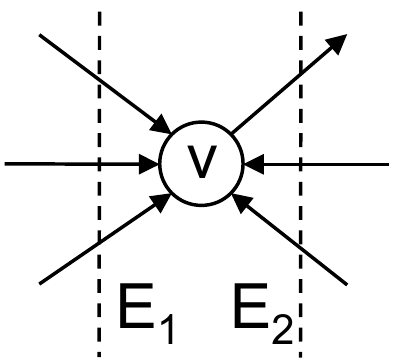}
        \caption{The inclusion of $v$ in $S$ or $T$ does not change the computed sum.}
        \label{fig:figs2_flip}
    \end{figure}

    Note that by switching a node between the two partitions does not change the computed area sum. That means that every $s-t$ cut correctly computes the value of the aggregate over area $G$, and since every edge corresponds to a unique datapoint, there is a direct correspondence between the minimum cut and the minimum number of data points that compute the spatial aggregate.\qed
\end{proof}

\begin{proof}[Proposition~\ref{prop:combined}]
    The result is straightforward. If a node is white in one case, and shaded in another, due to the fact that the source and target edges have infinite capacities the nodes cannot be in the same partition.\qed
\end{proof}

\begin{proof}[Proposition~\ref{prop:numOfCorners}]
    Dividing the region into rectangles uses existing corners and creates some new ones. New corners, i.e. points that were not corners in the original interest region but were created due to a division, will participate in the summation an even number of times, alternating between summation and addition, canceling each other out. This is made clear by Figure~\ref{fig:figs2_arbShape}. A ``created'' corner will have to participate into the sum an even number of times due to being a corner to two or four adjacent ``created'' rectangles. But as we can see from the right part of Figure~\ref{fig:figs2_arbShape}, which is a simple depiction of the way sums are computed for a simple rectangle, adjacent rectangles make the corner introduce its sum with alternating signs: positive and negative, depending on whether it is the lower right, lower left, upper right or upper left corner of the current rectangle.
    \begin{figure}[htbp]
      \centering
          \includegraphics[width=2.5in]{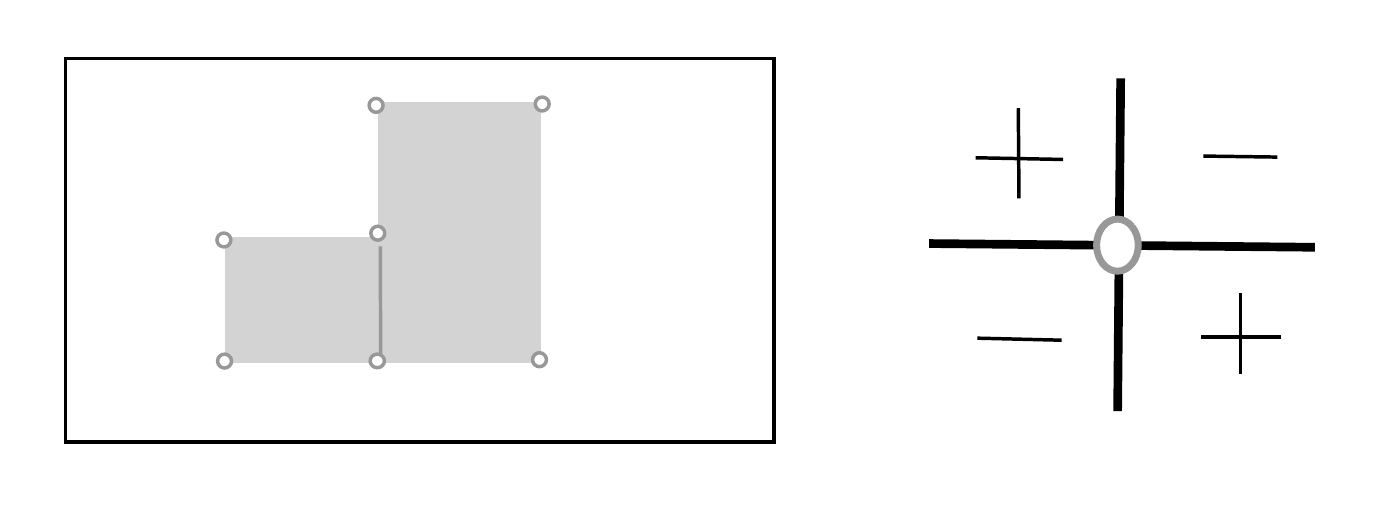}
      \caption{Each corner gets added or subtracted depending on its position relatively to the current rectangle.}
      \label{fig:figs2_arbShape}
      \vspace{3mm}
    \end{figure}
    
    The actual corners (the points that were corners in the original region) will participate to the summation once or thrice, ending up in the total either added once or subtracted once. Hence the result of the above lemma.
\end{proof}

\end{document}